\documentstyle[twocolumn,aps,epsf]{revtex}

\newcommand{\hbo}{\hbar \omega}
\newcommand{\om}{\omega}
\newcommand{\la}{\lambda}
\newcommand{\s}{\sigma}
\newcommand{\lm}{(\lambda,\mu)}
\newcommand{\lms}{(\lambda_{\sigma},\mu_{\sigma})}

\newcommand{\bea}{\begin{eqnarray}}
\newcommand{\eea}{ \end{eqnarray}}

\begin{document}

\draft

\wideabs{

\title{Partial Dynamical Symmetry in a Fermion System}
\author{Jutta Escher\cite{byline}
and Amiram Leviatan$^{**}$}
\address{Racah Institute of Physics, The Hebrew University,
Jerusalem 91904, Israel}
\date{\today}

\maketitle

\begin{abstract}
The relevance of the partial dynamical symmetry concept for
an interacting fermion system is demonstrated.
Hamiltonians with partial SU(3) symmetry are presented in
the framework of the symplectic shell-model of nuclei and shown to
be closely related to the quadrupole-quadrupole interaction.
Implications are discussed for the deformed light nucleus
$^{20}$Ne.
\end{abstract}

}

\narrowtext

Symmetries play an important role in dynamical systems.
They provide labels for the
classification of states, determine selection rules, and
simplify the relevant Hamiltonian matrices.
Algebraic, symmetry-based models offer significant simplifications
when the Hamiltonian under consideration commutes with all the generators
of a particular group (`exact symmetry') or when it is written in terms
of the Casimir operators of a chain of nested groups
(`dynamical symmetry') \cite{AlgTheo}. In both cases basis states belonging
to inequivalent irreducible representations
(irreps) of the relevant groups do not mix, the Hamiltonian matrix has block
structure, and all properties of the system can be expressed in closed form.
An exact or dynamical symmetry not only facilitates the numerical treatment
of the Hamiltonian, but also its interpretation and thus provides considerable
insight into the physics of a given system.

Naturally, the application of exact or dynamical symmetries to realistic
situations has its limitations.
Usually the assumed symmetry is only approximately fulfilled, and imposing
certain symmetry requirements on the Hamiltonian might result in constraints
which are too severe and incompatible with experimentally observed features
of the system.
The standard approach in such situations is to break the symmetry.
Partial Dynamical Symmetry (PDS) \cite{GenPDS} corresponds to a particular
symmetry-breaking for which the Hamiltonian is not invariant under the
symmetry group and hence various irreps are mixed in its eigenstates, yet it
possess a subset of `special' solvable states which respect the symmetry.
This new scheme has recently been introduced in bosonic systems and has
been applied to the spectroscopy of deformed nuclei \cite{Leviatan96a}
and to the study of mixed systems with coexisting regularity and
chaos \cite{PDSChaos}.
It is the purpose of this Letter to demonstrate the relevance of the partial
dynamical symmetry concept to fermion systems.
More specifically, in the framework of the symplectic shell-model of
nuclei \cite{SymplM},
we will prove the existence of a family of fermionic Hamiltonians with
partial SU(3) symmetry. The PDS Hamiltonians are rotationally invariant
and closely related to the quadrupole-quadrupole interaction;
hence our study will shed new light on this important interaction.
We will compare the spectra and eigenstates of the quadrupole-quadrupole
and PDS Hamiltonians for the deformed light nucleus $^{20}$Ne.

The quadrupole-quadrupole interaction is an important ingredient in models
that aim at reproducing quadrupole collective properties of nuclei.
A model which is able to fully accommodate the action of the collective
quadrupole operator,
$Q_{2m}=\sqrt{\frac{16\pi}{5}} \sum_s r^2_s Y_{2m} (\hat{r}_s)$,
is the symplectic shell model (SSM), an algebraic scheme which respects the
Pauli exclusion principle \cite{SymplM}.
In the SSM, this operator takes the form
$Q_{2m} = \sqrt{3} ( \hat{C}^{(11)}_{2m}
+ \hat{A}^{(20)}_{2m} + \hat{B}^{(02)}_{2m} )$,
where $\hat{A}^{(20)}_{lm},\,\hat{B}^{(02)}_{lm}$,
and $\hat{C}^{(11)}_{lm}$ are symplectic generators with good
SU(3) [superscript $(\lambda,\mu)$]
and SO(3) [subscript $l,m$] tensorial properties.
The $\hat{A}^{(20)}_{lm}$ ($\hat{B}^{(02)}_{lm}$), $l$ = 0 or 2,
create (annihilate) $2 \hbo$ excitations in the system.
The $\hat{C}^{(11)}_{lm}$, $l$ = 1 or 2, generate a SU(3) subgroup and
act only {\em within} one harmonic oscillator (h.o.\/) shell
($\sqrt{3} \hat{C}^{(11)}_{2m}=$ $Q^E_{2m}$, the symmetrized quadrupole
operator of Elliott, which does not couple different
h.o.\/ shells \cite{Elliott58}, and $\hat{C}^{(11)}_{1m}=\hat{L}_m$,
the orbital angular momentum operator).
A fermion realization of these generators is given in~\cite{Escher98b}.

\begin{figure}[htp]
\vspace{-6mm}
\hbox{
\hskip -10 pt
\epsfxsize=3.5 true in
\epsfbox{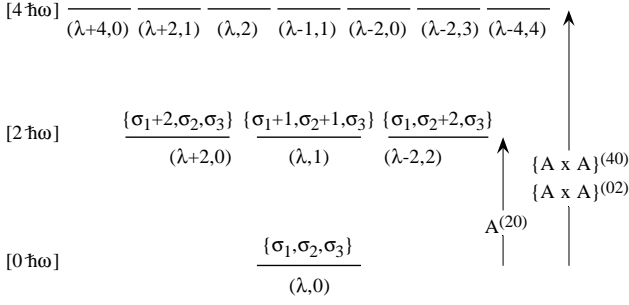}
}
\vskip 10 pt
\caption{
Basis construction in the symplectic model.
SU(3)-coupled products of the raising operator $\hat{A}^{(20)}$ with
itself act on an Elliott starting state with $\lms = (\la,0)$
($\{\s_1,\s_2,\s_3=\s_2\}$) to generate symplectic $2\hbo$, $4\hbo$,
$\ldots$ excitations.
Also shown are the SU(3) labels $\lm$ and quanta distributions
$\{\om_1,\om_2,\om_3\}$ for some excited states.
}
\label{SymplIrrep}
\end{figure}

A basis for the symplectic model is generated by applying symmetrically
coupled products of the 2$\hbo$ raising operator $\hat{A}^{(20)}$ with
itself to the usual $0 \hbo$ many-particle shell-model states.
Each $0 \hbo$ starting configuration
is characterized by the distribution of oscillator quanta into the three
cartesian directions, $\{ \s_1,\s_2,\s_3 \}$ ($\s_1 \geq \s_2 \geq \s_3$),
or, equivalently, by its U(1)$\times$SU(3) quantum numbers $N_{\s} \, \lms$.
Here $\la_{\s} = \s_1 - \s_2$,
$\mu_{\s} = \s_2 - \s_3$ are the Elliott SU(3) labels,
and $N_{\s} = \s_1 +\s_2 +\s_3$ is related to the eigenvalue of the
oscillator number operator.
The product of $N/2$ raising operators $\hat{A}^{(20)}$ generates $N\hbo$
excitations for each starting irrep $N_{\s} \, \lms$.
Each such product operator ${\cal P}^{N (\la_n,\mu_n)}$, labeled
according to its SU(3) content, $(\la_n,\mu_n)$, is coupled with
$| N_{\s} \, \lms \rangle$ to good SU(3) symmetry $\rho \lm$, with $\rho$
denoting the multiplicity of the coupling $(\la_n,\mu_n) \otimes \lms$.
The quanta distribution in the resulting state is given by
$\{ \om_1,\om_2,\om_3 \} $, with $N_{\s} + N = \om_1 + \om_2 + \om_3$,
$\om_1 \geq \om_2 \geq \om_3$, and $\la = \om_1 - \om_2$,
$\mu = \om_2 - \om_3$.
The basis state construction is schematically illustrated in
Fig.~\ref{SymplIrrep}
for a typical Elliott starting state with
$(\lambda_{\s},\mu_{\s}) = (\lambda, 0)$.
$^{20}$Ne, for instance, has $N_{\s}$ = 48.5 (after removal of the
center-of-mass contribution) and
$(\lambda_{\s},\mu_{\s})$ = (8,0) \cite{SymplM,Draayer84}.
To complete the basis state labeling, additional quantum num-\linebreak

\noindent
bers
$\alpha = \kappa L M$ are required, where $L$ denotes the angular momentum
with projection $M$, and $\kappa$ is a multiplicity index, which enumerates
multiple occurrences of a particular $L$ value in the SU(3) irrep $\lm$
from 1 to
$\kappa^{max}_L \lm = [ (\la+\mu+2-L)/2 ]$ - $[ (\la+1-L)/2 ]$
- $[ (\mu+1-L)/2 ]$, where [$\ldots$] is the greatest non-negative integer
function \cite{Lopez90}.
The group chain corresponding to this labeling scheme is Sp(6,R) $\supset$
SU(3) $\supset$ SO(3) which defines a dynamical symmetry basis.

The quadrupole-quadrupole interaction connects h.o.\/ states differing
in energy by $0\hbo$, $\pm 2\hbo$, and $\pm 4\hbo$, and may be written as
\bea
Q_2 \cdot Q_2 &=& 9 \hat{C}_{SU3} - 3 \hat{C}_{Sp6} +
\hat{H}_0^2 - 2 \hat{H}_0 - 3 \hat{L}^2
- 6 \hat{A}_0 \hat{B}_0 \nonumber \\
&& + \{ \mbox{terms coupling different h.o.\/ shells} \} \; , \label{Eq:QQ}
\eea
where  $\hat{C}_{SU3}$ and $\hat{C}_{Sp6}$ are the quadratic Casimir
invariants of SU(3) and Sp(6,R) with eigenvalues
$2(\la^2+\mu^2+\la\mu+3\la+3\mu)/3$ and
$2(\la_{\s}^2+\mu_{\s}^2+\la_{\s}\mu_{\s}+3\la_{\s}+3\mu_{\s})/3
+N_{\s}^2/3-4N_{\s}$, respectively.
These operators, as well as the h.o.\/ $\hat{H}_0$ and $\hat{L}^2$ terms,
are diagonal in the dynamical symmetry basis.
Unlike the Elliott quadrupole-quadrupole interaction,
$Q_2^E \cdot Q_2^E$ $=6 \hat{C}_{SU3} - 3 \hat{L}^2$, the
$Q_2 \cdot Q_2$ interaction of Eq.~(\ref{Eq:QQ}) breaks
SU(3) symmetry within each h.o.\/ shell since the term
$\hat{A}_0 \hat{B}_0 \equiv \hat{A}^{(20)}_0 \hat{B}^{(02)}_0 =
(\{ \hat{A} \times \hat{B} \}_0^{(00)}
- \sqrt{5} \{ \hat{A} \times \hat{B} \}_0^{(22)})/\sqrt{6}$
mixes different SU(3) irreps.
In order to study the action of $Q_2 \cdot Q_2$ within such a shell, we
consider the following family of Hamiltonians:
\bea
\lefteqn{H(\beta_0,\beta_2) = \beta_0 \hat{A}_0 \hat{B}_0
+ \beta_2 \hat{A}_2 \cdot \hat{B}_2 }
\label{Eq:Hpds} \\
&& = \frac{\beta_2}{18} ( 9\hat{C}_{SU3} - 9\hat{C}_{Sp6}
+ 3\hat{H}_0^2 - 36\hat{H}_0 )
   + ( \beta_0 - \beta_2) \hat{A}_0 \hat{B}_0 \; .  \nonumber
\eea
For $\beta_0=\beta_2$, one recovers the dynamical symmetry, and with
the special choice $\beta_0=12$, $\beta_2=18$, one obtains
$Q_2 \cdot Q_2 = H(\beta_0=12,\beta_2=18)+ const(N) - 3 \hat{L}^2$
+ terms coupling different shells, where $const(N)$ is constant for a
given h.o. $N\hbo$ excitation.

From Eq.~(\ref{Eq:Hpds}) it follows that $H(\beta_0,\beta_2)$ is not
SU(3) invariant. We will now show that $H(\beta_0,\beta_2)$ exhibits
partial SU(3) symmetry.
Specifically, we claim that among the eigenstates of $H(\beta_0,\beta_2)$,
there exists a subset of solvable pure-SU(3) states, the
SU(3)$\supset$SO(3) classification of which depends on both the Elliott
labels $(\lambda_{\s},\mu_{\s})$ of the starting state and the symplectic
excitation $N$.
In general, we find that all L-states in the starting configuration ($N=0$)
are solvable with good SU(3) symmetry $\lms$. For excited configurations
($N>0$ and even) we distinguish between two possible cases:
\begin{itemize}
\item[(a)] $\la_{\s} > \mu_{\s}$:
the pure states belong to $\lm = (\la_{\s}-N,\mu_{\s}+N)$ and have
$L = \mu_{\s}+N, \mu_{\s}+N+1, \ldots , \la_{\s}-N+1$
with $N=2,4, \ldots$ subject to $2N \leq (\la_{\s} - \mu_{\s} + 1)$.
\item[(b)] $\la_{\s} \leq \mu_{\s}$:
the special states belong to $\lm =(\la_{\s}+N,\mu_{\s})$ and have
$L = \la_{\s}+N, \la_{\s}+N+1, \ldots , \la_{\s}+N+\mu_{\s}$
with $N=2,4, \ldots$.
\end{itemize}

To prove the claim, it suffices to show that
$\hat{B}_0$ annihilates the states in question.
For $N=0$ this follows immediately from the fact that the $0 \hbo$
starting configuration is a Sp(6,R) lowest weight which, by definition,
is annihilated by the lowering operators of the Sp(6,R) algebra.
The latter include the generators $\hat{B}^{(02)}_{lm}$. For $N>0$,
let $\{ \sigma_1,\sigma_2,\sigma_3 \}$ be the quanta distribution for a
0$\hbo$ state with $\la_{\s} > \mu_{\s}$.
Adding $N$ quanta to the 2-direction yields a $N \hbo$ state
with quanta distribution $\{ \s_1, \s_2+N, \s_3 \}$, that is
$\lm = (\la_{\s}-N,\mu_{\s}+N)$.
Acting with the rotational invariant $\hat{B}_0$ on such a state does not
affect the angular momentum, but removes two quanta from
the 2-direction, giving a $(N-2) \hbo$ state with
$(\la',\mu') = (\la_{\s}-N+2,\mu_{\s}+N-2)$.
(The symplectic generator $\hat{B}_0$ cannot remove quanta from the other two
directions of this particular state, since this would yield a state belonging
to a different symplectic irrep.)
Comparing the number of $L$ occurrences in $\lm$ and $(\la',\mu')$, one finds
that as long as $\la_{\s} - N + 1 \geq \mu_{\s}+N$,
$\Delta_L(N) \equiv \kappa_L^{max}\lm - \kappa_L^{max}(\la',\mu') = 1$
for $L = \mu_{\s}+N, \mu_{\s}+N+1, \ldots, \la_{\s}-N+1$, and
$\Delta_L(N)=0$ otherwise.
When $\Delta_L(N)$=1, a linear combination $|\phi_L(N)\rangle =
\sum_{\kappa} c_{\kappa} | N\hbo (\la_{\s}-N,\mu_{\s}+N) \kappa L M \rangle$
exists such that $\hat{B}_0 |\phi_L(N)\rangle = 0$, and thus our claim
for family (a) holds. The proof for family (b) can be carried out analogously
if one considers adding $N$ quanta to the 1-direction of the starting irrep.
In this case there is no restriction on N, hence family (b) is infinite.

The special states have well defined symmetry
Sp(6,R) $\supset$ SU(3) $\supset$ SO(3) and are annihilated by $\hat{B}_0$.
This ensures that they are solvable eigenstates of $H(\beta_0,\beta_2)$
with eigenvalues $E(N=0) = 0$,
$E(N) = \beta_2 N( N_{\s}-\la_{\s}+\mu_{\s}-6+3N/2)/3$ for
family (a), and $E(N) =  \beta_2 N( N_{\s}+2\la_{\s}+\mu_{\s}- 3 + 3N/2 )/3$
for family (b). All 0$\hbo$ states are unmixed and span the entire $\lms$
irrep. In contrast, for the excited levels ($N > 0$),
the pure states span only part of the corresponding SU(3)
irreps.
There are other states at each excited level which do not preserve the
SU(3) symmetry and therefore con-
 \linebreak

\begin{figure}
\begin{center}
\leavevmode
\hbox{%
\hskip -7 pt
\epsfysize=1.9 true in
\epsfbox{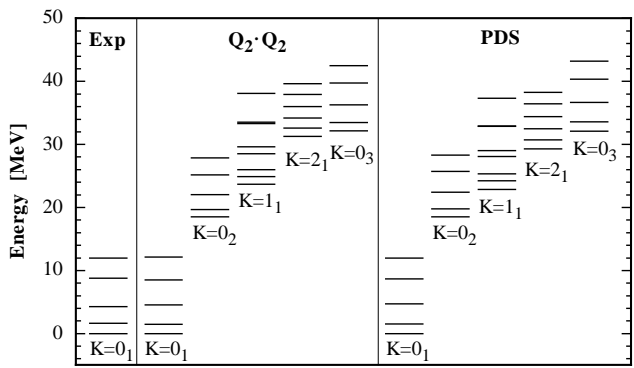}}
\end{center}
\vspace{-1mm}
\caption{
Energy spectra for $^{20}$Ne.
Comparison between experimental values (left), results from a symplectic
$8\hbo$ calculation (center) and a PDS calculation (right).
The angular momenta of the positive parity states in the rotational bands are
$L$=0,2,4,$\ldots$ for K=0 and $L$=K,K+1,K+2, $\ldots$ otherwise.
}
\label{Energies}
\end{figure}

\noindent
tain a mixture of SU(3) irreps.
The partial SU(3) symmetry of $H(\beta_0,\beta_2)$ is converted into partial
dynamical SU(3) symmetry by adding to it SO(3) rotation terms which lead
to L(L+1)-type splitting but do not affect the wave functions.
The solvable states then form rotational bands and since their wave
functions are known, one can evaluate the E2 rates between them
\cite{EscherPDS2}. It is of interest to note that both the fermion
Hamiltonian presented here and the boson Hamiltonian of \cite{Leviatan96a}
exhibit partial SU(3) symmetry and involve a SU(3) tensor of the form
$[(2,0)\times (0,2)](2,2)L=0$.

To illustrate that the PDS Hamiltonians of Eq.~(\ref{Eq:Hpds}) are
physically relevant, we compare the eigenstates of $H_{PDS} = h(N) +
\xi H(\beta_{0}=12,\beta_{2}=18) + \gamma_2 \hat{L}^2 + \gamma_4
\hat{L}^4$ to those of the symplectic Hamiltonian $H_{Sp6} = \hat{H}_0
- \chi Q_2 \cdot Q_2 + d_2 \hat{L}^2 + d_4 \hat{L}^4$.  Here the
function $h(N)$ is simply a constant for a given $N\hbo$ excitation
and contains the h.o.\/ term $\hat{H}_0$.  Least squares fits to
measured energies and B(E2) values of the ground band of $^{20}$Ne
were carried out for 2$\hbo$, 4$\hbo$, 6$\hbo$, and 8$\hbo$ symplectic
model spaces.The resulting energies and transition rates converge to
values which agree with the data, Fig.~\ref{Energies} and
Table~\ref{Tab:Ne20Be2}. The parameters $\gamma_2$ and $\gamma_4$ in
$H_{PDS}$ were determined by the energy splitting between states of
the ground band, $\xi$ was adjusted to reproduce the relative

\begin{table}[hbtp]
\vspace{-1mm}
\caption{B(E2) values (in Weisskopf units)
for ground band transitions in $^{20}$Ne.
Compared are several symplectic calculations, PDS results,
and experimental data \protect\cite{Tilley98}.
The static quadrupole moment of the $2^+_1$ state is given in the last row.
PDS results are rescaled by an effective charge e$^*$=1.95 and
the symplectic calculations employ bare charges.}
\vspace{-2mm}
\begin{center}
\begin{tabular}{ccccccccc}
\multicolumn{1}{c}{Transition} & \multicolumn{5}{c}{Model B(E2) [W.u.]} &
B(E2) [W.u.] \\
 \cline{2-6}
 $J_i \rightarrow J_f$ & \multicolumn{1}{c}{$2\hbar\om$} &
 \multicolumn{1}{c}{$4\hbar\om$} & \multicolumn{1}{c}{$6\hbar\om$} &
 \multicolumn{1}{c}{$8\hbar\om$} & \multicolumn{1}{c}{PDS} & Exp. \\
\hline
  2 $\rightarrow$ 0 & 14.0 & 18.7 & 19.1 & 19.3 & 20.3 &  20.3 $\pm$ 1.0 \\
  4 $\rightarrow$ 2 & 18.4 & 24.5 & 24.6 & 24.5 & 25.7 &  22.0 $\pm$ 2.0 \\
  6 $\rightarrow$ 4 & 17.1 & 22.3 & 21.5 & 20.9 & 21.8 &  20.0 $\pm$ 3.0 \\
  8 $\rightarrow$ 6 & 12.4 & 15.2 & 13.3 & 12.4 & 12.9 &   9.0 $\pm$ 1.3 \\
\hline
\multicolumn{1}{c}{$Q$ [eb]}
    & -0.14 & -0.16 & -0.16 & -0.16 & -0.17 & -0.23 $\pm$ 0.03 \\
\end{tabular}
\end{center}
\vspace{0.1in}
\label{Tab:Ne20Be2}
\end{table}                                                

\noindent
positions of the resonance
bandheads and $h(N)$ was fixed by the energy difference
$[E(0_2^+) - E(0_1^+)]$.
Fig.~\ref{Energies} and Table~\ref{Tab:Ne20Be2} demonstrate the level
of agreement between the PDS and symplectic results.

\begin{figure}
\hbox{
\hskip -22 pt
\epsfysize=7.6 true in
\epsfbox{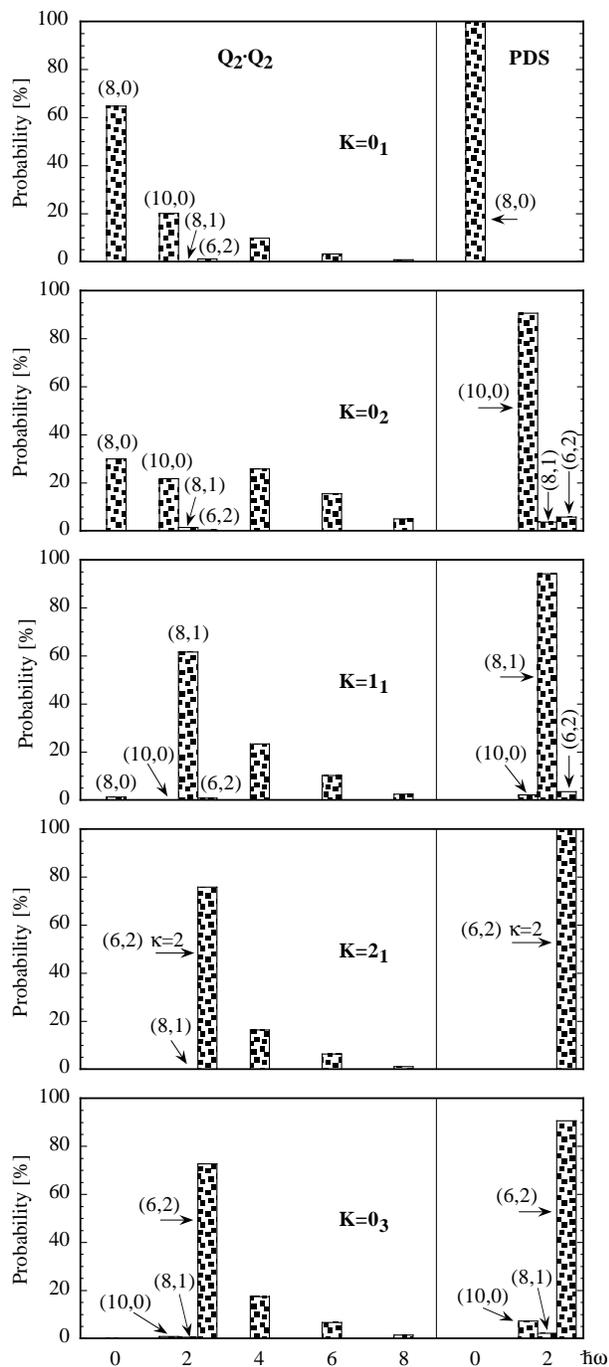}
}
\caption{
Decomposition for calculated $2^+$ states of $^{20}$Ne.
Individual contributions from the relevant SU(3) irreps at the 0$\hbo$
and 2$\hbo$ levels are shown for both a symplectic $8\hbo$ calculation
(denoted $Q_{2}\cdot Q_{2}$) and a PDS calculation.
In addition, the total strengths contributed by the $N\hbo$ excitations
for $N>2$ are given for the symplectic case.
}
\label{Decomp}
\end{figure}

An analysis of the structure of the ground and resonance bands reveals
the amount of mixing in the 8$\hbo$ symplectic ($Q_{2}\cdot Q_{2}$)
wave functions.
Fig.~\ref{Decomp} shows the decomposition for representative ($2^{+}$)
states of the five lowest rotational bands.
Ground band (K=$0_1$) states are found to have a strong $0\hbo$ component
($\geq 64\%$), and three of the four resonance bands are clearly
dominated ($\geq 60\%$) by $2\hbo$ configurations.
States of the first resonance band (K=$0_2$), however, contain significant
contributions from all but the highest $N\hbo$ excitations.
The relative strengths of the SU(3) irreps within the $2\hbo$ space are
shown as well:  states are found to be dominated by one representation
[(10,0) for the K=$0_2$ band, (8,1) for K=$1_1$, (6,2)$\kappa=2$ for
K=$2_1$, and (6,2)$\kappa=1$ for K=$0_3$, where $\kappa=1$ and 2 correspond
here to Vergados basis labels 0 and 2, respectively \cite{Vergados}],
while the other irreps contribute
only a few percent. Such trends are present also in the more realistic
symplectic calculations of \cite{Suzuki87}.

The PDS Hamiltonian $H_{PDS}$ acts only within one oscillator shell,
hence its eigenfunctions do not contain admixtures from different
$N\hbo$ configurations.
As expected, $H_{PDS}$ has families of pure SU(3) eigenstates which can be
organized into rotational bands.
The ground band belongs entirely to $N=0$, $\lm=(8,0)$, and all states of
the K=$2_1$ band have quantum labels $N=2$, $\lm=(6,2)$, $\kappa=2$.
A comparison with the symplectic case shows that the $N\hbo$ level to
which a particular PDS band belongs is also dominant in the corresponding
symplectic band.
In addition, within this dominant excitation, eigenstates of $H_{PDS}$
and $H_{Sp6}$ have similar SU(3) distributions; in particular, both
Hamiltonians favor the same $\lm\kappa$ values.
Significant differences in the structure of the wave functions appear,
however, for the K=$0_2$ resonance band.
In the $8 \hbo$ symplectic calculation, this band contains almost equal
contributions from the $0 \hbo$, $2 \hbo$, and $4 \hbo$ levels, with
additional admixtures of $6 \hbo$ and $8 \hbo$ configurations, while in
the PDS calculation, it belongs entirely to the $2 \hbo$ level.
These structural differences are also evident in the interband transition rates, 
e.g.\/ B(E2; K=$0_1$, L=$2^+$ $\rightarrow$ K=$0_2$, L=$0^+$) = 2.93 (5.69) W.u.\/
and B(E2; K=$0_2$, L=$2^+$ $\rightarrow$ K=$0_1$, L=$0^+$) = 5.84 (12.6) W.u.\/
in the $8 \hbo$ (PDS) calculation, and reflect the action of the inter-shell
coupling terms in Eq.~(\ref{Eq:QQ}).  
Increasing the strength $\chi$ of $Q_2 \cdot Q_2$ in $H_{Sp6}$
will also spread the other resonance bands over many $N\hbo$ excitations.
The K=$2_1$ band (which is pure in the PDS scheme) is found to resist
this spreading more strongly than the other resonances.
For physically relevant values of $\chi$, the low-lying bands have
the structure shown in Fig.~\ref{Decomp}.

In summary, we have introduced a family of {\em fermionic} Hamiltonians
with partial SU(3) symmetry.
Using the framework of the symplectic shell model, we have proven

\noindent
 that
these Hamiltonians possess both mixed-symmetry and solvable
pure-SU(3) rotational bands.
For the deformed light nucleus $^{20}$Ne, we have shown
that various features of the quadrupole-quadrupole
interaction can be reproduced with a particular parameterization of
the PDS Hamiltonians.
For both the ground and the resonance bands, PDS eigenstates were seen
to approximately reproduce the structure of the exact $Q_2 \cdot Q_2$
eigenstates within the $0 \hbo$ and $2 \hbo$ spaces, respectively.
In particular, for each pure state of the PDS scheme we found a
corresponding eigenstate of the quadrupole-quadrupole interaction,
which was dominated by the same SU(3) irrep.
Moreover, for reasonable interaction parameters, each rotational band
was primarily located in one level of excitation, with the exception of
the lowest K=$0_2$ resonance band, which was spread over many
$N\hbo$ excitations.
Implications of the structural differences between the various resonance
bands for giant monopole and quadrupole transitions remain to be investigated.
The occurrence of partial symmetries for fermions, as shown in this work,
and for bosons, as presented in previous works \cite{Leviatan96a},
highlights their relevance to dynamical systems and motivates their
further study.

The authors acknowledge valuable suggestions by D.\/J.\/ Rowe
and constructive comments by J.\/P.\/ Elliott and G.\/ Rosensteel.
This work is supported by a grant from the
Israel Science Foundation. We thank the Institute for Nuclear Theory at the
University of Washington for its hospitality and the Department of Energy for
partial support during the completion of this work.

\end{document}